# Enhancing MBSE Education with Version Control and Automated Feedback

Levente Bajczi[1], Dániel Szekeres[1], Daniel Siegl[2], Vince Molnár[1]

[1] Budapest University of Technology and Economics
Department of Artificial Intelligence and Systems Engineering

[2] LieberLieber Software GmbH

**Abstract**: This paper presents an innovative approach to conducting a Model-Based Systems Engineering (MBSE) course, engaging over 80 participants annually. The course is structured around collaborative group assignments, where students utilize Enterprise Architect to complete complex systems engineering tasks across six submissions. This year, we introduced several technological advancements to enhance the learning experience, including the use of LemonTree, SmartGit, and GitHub. Students collaborated on shared repositories in GitHub, received continuous feedback via automated checks through LemonTree Automation, and documented their progress with pre-rendered, continuously updating diagrams. Additionally, they managed 2-way and 3-way merges directly in SmartGit, with merge issues, updates, and model statistics readily available for each Work-in-Progress submission. The process of correcting and providing manual feedback was streamlined, thanks to accessible changelogs and renders in GitHub. An end-of-course feedback form revealed high student satisfaction.

**Acknowledgement**: We would like to thank LieberLieber Software GmbH, Sparx Systems Pty Ltd., and syntevo GmbH for providing educational licenses for our course.

## 1 Introduction

Model-Based Systems Engineering (MBSE) has become a cornerstone in modern engineering practices. With the increasing intricacies of system design, MBSE offers a structured methodology that enhances clarity, traceability, and communication among stakeholders. As industries shift towards more sophisticated and interconnected systems, the demand for engineers proficient in MBSE methodologies and tools is growing [1]. This necessitates an educational paradigm that equips students with both theoretical knowledge and practical experience in using state-of-the-art tools.

In the current industry landscape, the integration of automated reviews and version control has become essential [3,4]. Automated reviews serve as a critical first step, catching errors and inconsistencies early in the development process. This not only enhances the quality of the systems being developed but also significantly reduces the time and effort required for manual reviews. Similarly, version control systems like Git are crucial for tracking changes, facilitating collaboration, and maintaining a history of model evolution. By ensuring that students are familiar with these tools, we prepare them to enter the workforce with the skills needed to streamline and optimize engineering workflows.

Our MBSE course teaches systems engineering with practical examples in SysML [2] to over 80 students yearly. This academic year, we innovated our practical homework assignment by leveraging tools such as LemonTree, SmartGit, and GitHub, aiming to bridge the gap between academic learning and industry requirements [7], help students make better models, and lessen the workload on reviewers. By integrating these tools into the curriculum, we provide students with hands-on experience in a collaborative and automated environment, mirroring the practices they will encounter in their professional careers. This effort is orthogonal to previous reports on MBSE education such as [1,5,6], and represents a collaboration between academia and tool vendors [7], driven by needs from industrial partners. This paper outlines the structure, implementation, and outcomes of this innovative approach, demonstrating its effectiveness in enhancing the learning experience.

### 1.1 Toolchain Details

**Enterprise Architect** Enterprise Architect serves as the cornerstone of our MBSE course, providing a robust platform for students to model and design complex systems. The tool supports various modeling languages including SysML, enabling students to capture and analyze system requirements, architecture, and behavior effectively. By using Enterprise Architect, students gain hands-on experience with an industry-standard tool.

**GitHub** GitHub plays a critical role in facilitating collaboration and version control in our MBSE course. Each student group works on shared repositories hosted on GitHub, which allows them to manage their project files efficiently and track changes over time. GitHub's platform supports the opening and merging of pull requests (PRs), enabling automated workflows for assignment submissions. This collaborative environment not only mirrors professional development practices but also enhances students' understanding of version control and collaborative project management.

**GitHub Actions** GitHub Actions is a key component of the course infrastructure, enabling automated workflows that streamline the development and feedback process. With GitHub Actions, we have set up automated checks that trigger with each pull request submission, providing continuous integration and immediate feedback on students' work. These automated workflows include running LemonTree Automation to validate model changes. This not only enhances the efficiency of the teaching process but also gives students a real-world understanding of continuous integration and delivery (CI/CD) practices, preparing them for professional environments where such automation is becoming standard.

**SmartGit** SmartGit is an efficient tool for handling the complexities of version control and merging in our course. Students use SmartGit to manage 2-way and 3-way merges directly, resolving conflicts that arise during collaborative work. This capability is particularly important for maintaining the integrity of shared models and ensuring seamless integration of individual contributions. By working with SmartGit, students learn valuable skills in conflict resolution and efficient version control, which are crucial for managing large-scale engineering projects in the real world.

**LemonTree** LemonTree significantly enhances our course by providing graphical diff/merge capabilities, as well as automated checks and continuous feedback on students' progress. It analyzes model changes and ensures consistency and compliance with predefined standards, giving students immediate insights into their work. Additionally, LemonTree Automation supports the generation of diagram renders, which are crucial for maintaining up-to-date documentation. By integrating it into our workflow, we streamline the feedback process and ensure that students can focus on refining their models with confidence. We used LemonTree Automation in the back-end of our GitHub Actions infrastructure, while students performed model merges and conflict resolution via LemonTree Desktop's integration with SmartGit.

## 1.2 Course Details

**Previous Knowledge Expected from Students** Students enrolling in the MBSE course are expected to have a foundational understanding of model-based engineering principles (such as UML in the context of software engineering) and basic familiarity with modeling tools. Prior experience with software development practices, including version control systems like Git, is also beneficial, but the tutorial warm-up exercise covers these details. While the practical parts of the course provide a tutorial on merging and conflict resolution, a basic competence in handling repositories and collaborative workflows is beneficial.

**Course Outline** The course covers a comprehensive range of topics integral to MBSE. While the theoretical part of the course tries to remain as generic as possible, the practical part (the 6-step homework assignment) covers the following key topics specifically in SysML:

1. Requirement analysis
2. Structural modeling
3. Fault tolerance
4. Behavior modeling
5. Platform modeling
6. Verification & Validation

Except for the Fault tolerance step, all these topics were expected to be done in Enterprise Architect, using the infrastructure presented in this paper.

**Expected Learning Outcomes** Upon completing the course, students are expected to achieve the following outcomes:

- Proficiency in MBSE Tools: Demonstrate competence in using Enterprise Architect, LemonTree, SmartGit, and GitHub for collaborative model-based systems engineering tasks.
- Enhanced Collaboration Skills: Effectively work in teams, managing shared repositories and resolving conflicts to maintain model integrity.
- Automated Review Processes: Interface with automated checks and continuous feedback mechanisms to improve the quality and consistency of models.
- Comprehensive Documentation: Produce and maintain accurate, up-to-date documentation that reflects ongoing model changes and updates.

- Industry Readiness: Gain practical experience with industrial tools and processes, preparing for seamless integration into professional engineering environments.

## 2 The Student Workflow

In our MBSE course, the student workflow is designed to mimic professional engineering practices, utilizing automated processes to streamline assignments and submissions. The overview of the process is shown in Figure 1.

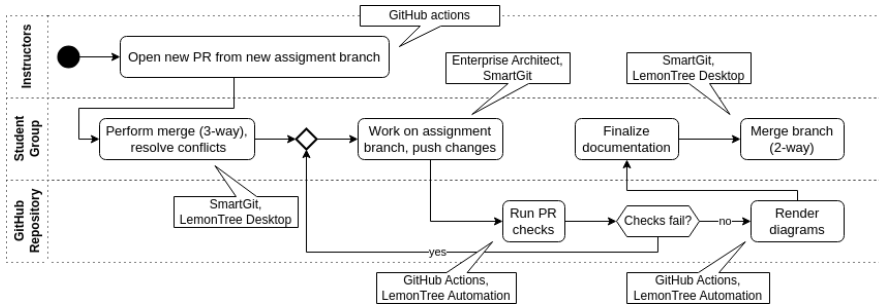

Figure 1. Overview of the student workflow

### 2.1 Assignment is Automated PR Opening

The workflow begins with the automated opening of pull requests (PRs) on GitHub for each group assignment. At the start of each assignment cycle, an automated process creates a new branch in the group's shared repository. This branch is where students will develop their models and complete their tasks. The automation also generates an initial pull request, setting up a structured framework for tracking changes and facilitating continuous integration.

This automation ensures consistency and provides a clear starting point for each assignment. It helps students focus on their engineering tasks without worrying about the intricacies of setting up the version control environment. Additionally, it standardizes the process, making it easier for instructors to manage and monitor progress across all groups.

### 2.2 Submission is PR Merging

Students first need to perform a 3-way merge among their previous changes, the new handout's changes, and the original model. Then, once students have completed their assignment, the submission process involves merging the PR into the main branch of their repository. These steps are facilitated by SmartGit and LemonTree, which students use to handle any model-based merge conflicts that may arise. Before merging, the PR undergoes a series automated checks via GitHub Actions and LemonTree Automation, validating the model's quality, consistency, and compliance with requirements.

The results of these checks are provided as feedback within the PR, allowing students to make necessary adjustments before the final merge. Once the checks are passed and any conflicts are resolved, students can proceed to merge the PR, completing the submission process. This workflow ensures that the models are thoroughly vetted and gives students immediate feedback on their work, helping them learn and improve continuously. Using "required" checks on the pull requests, we could ensure that students may only merge ready-to-submit models for their assignments.

**2.3. Documentation is Markdown with Dynamically Updating Diagram Renders**

In our course, documentation plays a crucial role in capturing the design decisions, requirements, and evolution of the models created by the students. To streamline this process, we utilize Markdown and dynamically updating diagram renders for writing documentation. This offers several advantages in terms of accessibility, readability, and real-time updates.

Markdown is a lightweight markup language that allows students to create well-structured and easy-to-read documentation containing clear descriptions of their models, requirements, and design processes. Markdown files are stored in the same GitHub repository as the models, ensuring that documentation and models are kept together and version-controlled.

The integration of dynamically updating diagram renders is facilitated through automated workflows using GitHub Actions and LemonTree Automation. Whenever students make changes to their model and commit them to the repository, the diagrams are automatically updated and rendered. These updated diagrams are then embedded directly into the Markdown files, providing an up-to-date visual representation of the models. This dynamic documentation process ensures that the diagrams always reflect the latest state of the models, eliminating the need for manual updates and reducing the risk of inconsistencies between the documentation and the actual models. It also allows instructors to review the models more effectively, as they can see both the textual descriptions and the corresponding visual representations in one place, without opening the modeling tool.

By leveraging Markdown with dynamically updating diagram renders, students learn to maintain accurate and up-to-date documentation throughout the project lifecycle. This mirrors industry standards where documentation must be continuously updated to reflect the current state of the system design. It also improves the learning outcomes by emphasizing the importance of clear and comprehensive documentation in systems engineering.

# 3 Implementation

This section details the usage of the previously introduced tools to achieve our objectives, including the management of course materials and student submissions, the automation of feedback, and the continuous integration of documentation.

### 3.1 Course Management via GitHub

We chose GitHub as the primary platform for managing student submissions due to its robust version control features and widespread familiarity among software developers. We opted not to use GitHub Classroom for a few reasons, mainly revolving around the need to maintain the history of the template repository in student repositories and to facilitate pull requests.

**Template Repository and Forking** Instead of using GitHub Classroom, which creates isolated repositories for each student without retaining the template repository's history, we created a central template repository and made forks for student groups. This approach preserved the entire commit history of the template in student repositories, allowing us to use pull requests (PRs) effectively. This was crucial for our workflow, which relied on students submitting their work via pre-opened PRs.

**Repository Structure and Access Control** Each student group was provided with a private fork of the central template repository. This structure allowed for:

- Independent Workspaces: Each group had its own private repository, ensuring their work remained isolated from others.
- Centralized Feedback: By using the pull request mechanism, we could centralize feedback and discussions related to specific submissions directly on GitHub, maintaining a clear and organized record of interactions.
- Access control via GitHub's built-in team and permission settings: This ensured that only authorized personnel (instructors and teaching assistants) had the necessary access to student repositories for review and feedback.

### 3.2 The Tutorial

To prepare students to work with Enterprise Architect, SmartGit, and LemonTree, we began the course with an optional tutorial. This tutorial was designed to guide them through the installation, configuration, and basic usage of these tools.

**Installation and Licensing** The tutorial began with detailed instructions for installing and licensing the required software. Each step was clearly outlined, including the necessary files, activation codes, and server credentials.

- Enterprise Architect: Students were instructed to download and run the installer, following the wizard to complete the installation. Then, they were guided through adding a license key via a shared keystore, using the provided server details to test and apply the key. A final activation step involved entering an activation code.
- SmartGit: Installation steps were similarly straightforward, requiring administrative permissions and default settings. Post-installation, students registered their SmartGit licenses and configured the software to use their standard GitHub credentials. Instructions were provided for generating and adding SSH keys to GitHub for secure authentication.
- LemonTree Desktop: The installer was run with specific instructions for SmartGit integration. Students then applied their LemonTree license using the provided file.

**Basic Tasks and Workflow Familiarization** The tutorial then moved on to practical tasks to familiarize students with the essential functionalities of the tools. Students opened a sample model in Enterprise Architect and made a simple modification by renaming a class. Then, they committed these changes to a new branch in SmartGit, using LemonTree to visualize and inspect the changes before committing. They were instructed to push their changes to GitHub and open a PR from their branch to the main branch. The tutorial explained how to monitor the automated checks triggered by the PR and interpret the results.

To introduce merging, students repeated the modification task on a different branch, leading to a merge conflict. The tutorial provided a step-by-step guide on using SmartGit and LemonTree to resolve these conflicts. LemonTree's three-way merge visualization was highlighted as a key tool for understanding and resolving discrepancies between branches. Students were then shown how to commit the merged changes and push them to GitHub. They were instructed to merge the PR, completing the cycle from modification to integration.

### 3.3 Automated Checks for WiP Submissions

To enhance the learning experience through continuous feedback, we integrated a series of automated checks for students' work-in-progress (WiP) submissions using GitHub Actions. These checks were designed to ensure the quality, consistency, and correctness of the models students submitted.

The automated checks included a "diff check", which compared changes between the base and head branches of a pull request, allowing students and instructors to visualize and review modifications. A "model check" was also implemented to validate the integrity and structure of the submitted models, ensuring they adhered to some required standards. Additionally, a "consistency check" was performed to detect and report any inconsistencies within the models, further maintaining the quality of the work.

These automated processes provided timely feedback directly within the GitHub pull request, enabling students to address issues promptly and improving the overall workflow.

## 4 Evaluation

The infrastructure's effectiveness was evaluated qualitatively through the feedback of the graders regarding both the infrastructure and the quality of submissions, and student feedback on the overall experience.

### 4.1 Grader feedback

**Regarding the infrastructure** Having the automated renders inside the documentation made it much faster to evaluate the submissions. There were cases where it was necessary to open the model itself because of some rendering issue, but it was still better than always relying on manual exports done by the students or opening the model. The automated consistency checks filtered out mistakes in the submissions which we previously had to check one by one, allowing the graders to focus on the design decisions and semantic correctness.

**Regarding the quality of submissions** The model-aware version control and three-way merge option allowed student groups to assign individual tasks to group members and work on the homework in parallel much more easily. The better time management and having early feedback from the automated checks made the quality of the submissions higher than before.

### 4.2 Student Feedback

At the end of the course, students were asked to rate their experience with the new infrastructure (see Figure 2). The feedback indicated a relatively high level of satisfaction, with more than 70% of the responses being neutral or positive. This can be considered a very good result since the students never had any experience with similar technology before. Therefore, the score on this question shall not be interpreted as a comparison to other approaches but rather as a general impression of comfort and usability, where a neutral answer indicates the lack of serious issues or complications.

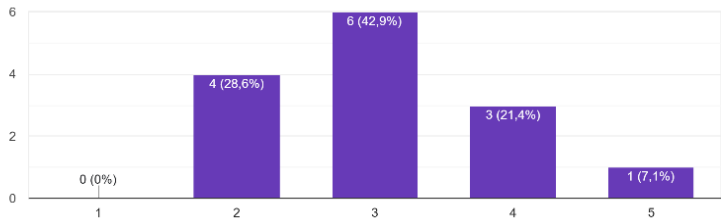

Figure 2. Distribution of answers on a 5-point Likert scale to "How satisfied were you with the whole infrastructure (Enterprise Architect, LemonTree, Github)?" (1 – "I never want to use it again"; 5 – "I think it was excellent")

Informal feedback showed that students appreciated LemonTree as a standalone tool but had a hard time working with merge conflicts in general. Based on feedback from previous years, where the typical solution for merge conflicts was to discard one of the conflicting versions, this can be considered as a great improvement. One student indicated that model merging should be emphasized in the course material and the optional tutorial should be mandatory.

Overall, the new infrastructure was successful in enhancing both the learning experience and the quality of the submissions, reflected in the completion rate (79%) and positive feedback.

## 5 Summary and Future Work

The integration of GitHub, LemonTree, SmartGit, and GitHub Actions into our Model-Based Systems Engineering (MBSE) course significantly improved the efficiency and quality of the student experience. By automating the submission and feedback process, we were able to provide continuous, actionable insights that enhanced students' understanding and engagement. The new infrastructure facilitated smoother collaboration, effective conflict resolution, and timely feedback, all of which contributed to higher completion rates and improved quality of submissions. The evaluation of the new infrastructure demonstrated its success with a significant number of groups completing assignments to a satisfactory or exceptional degree. Measuring student satisfaction showed mostly neutral and positive responses, which indicates a smooth and comfortable experience for the students.

Looking forward, we aim to further enhance our infrastructure by integrating a pattern-based model validator. This advanced tool will help automate the validation of the semantic correctness of the submissions, ensuring that the models not only meet structural requirements but also adhere to best practices in system design. The pattern-based validator will provide deeper insights into the quality and functionality of the models, allowing students to receive more comprehensive feedback on their work. By continuing to innovate and integrate new technologies into our course, we strive to maintain a cutting-edge educational experience that prepares our students for the complexities of real-world systems engineering challenges.